\documentclass{elsart}

\usepackage{natbib}

\usepackage{psfig}

\usepackage{amssymb}

\begin{document}

\begin{frontmatter}



\title{Jets from subpc to kpc scale}


\author{Fabrizio Tavecchio}

\address{Osservatorio Astronomico di Brera-Merate, Via Bianchi 46,
Merate-Italy}

\begin{abstract}
The {\it Chandra} discovery of bright X-ray emission from kpc-scale jets
provides us unprecedented insights into the physical state of the plasma
in the flow. In particular it is possible to get good constraints on the
power and pressure in bright knots.  For a group of selected
sources with blazar-type cores it is also possible to constrain the
physical quantities of the jet at sub-pc scale. We discuss how these
results can help us to connect the properties of the jet at different
scales.
\end{abstract}


\end{frontmatter}

\section{Introduction}
\label{intro}
The study of extragalactic jets has been recently renewed by the
discovery, made by {\it Chandra}, of numerous jets bright in the X-rays
at scales $>$ kpc, both in FRI and FRII sources (Chartas et al. 2000,
Sambruna et al. 2002). While in low-power FRI jets the dominating
emission mechanism responsible for X-rays is thought to be synchrotron
from very-high energy electrons (e.g. Worrall et al. 2001), in the case
of the most powerful sources it is clear that X-rays constitute a second
spectral component, likely due to the IC scattering of the CMB photons
(Tavecchio et al. 2000a, Celotti et al. 2001), implying that X-ray bright
jets are still relativistic (with bulk Lorentz factors $5-10$) at $\sim
100$ kpc.\\ At the opposite side of scalelengths, studies of the
innermost portion ($d\sim 0.1$ pc) of jets in AGNs are rather well
developed.  Our knowledge of these regions of jets is mainly based on
blazars, whose Spectral Energy Distribution (extending from radio to
$\gamma$-rays) is dominated by the relativistic amplified nonthermal
continuum produced in a jet closely aligned to the line of sight
(e.g. Urry \& Padovani 1995). The double humped SED of these sources is
well described by synchrotron-IC models (e.g., Ghisellini et al. 1998;
Sikora \& Madejski 2001).\\ The possibility to constrain the physical
state of the plasma in the jet both at sub-pc and kpc scale offers us the
interesting opportunity to shed some light on the evolution of the jet
from very small scales, close to the central engine, to the outer
regions, where the jet is starting to significantly decelerate. In this
work we present a first step, reporting the results obtained for a small
group of sources for which good data for both regions are
available. These sources are PKS~1510-089, 1641+399, PKS~0521-365 and
3C371. The emission models from the blazar region and the parameters for
the first three sources are discussed in Tavecchio et al. (2000b, 2002),
while the model for 3C371 is discussed in Tavecchio et al. (in
prep). Multiwavelength ({\it Chandra}, {\it HST} and radio) data for
extended jets are discussed in Gambill et al. (in prep 1510-089;
1641+399), Pesce et al. (2001, 3C371) and Birkinshaw et al. (2002,
0521-365). Clearly the X-ray emission for the first two sources is well
reproduced by the IC/CMB model, while in the other two jets the dominant
contribution is due to synchrotron emission. The detailed modelling of
the data will be reported in Tavecchio et al. (in prep). In the following
we limit the discussion to the result of the comparison of the most
relevant quantities of the jet as inferred in the two regions.

\section{Results: power and pressure}

A fundamental global quantity of jets is the total (particles + magnetic
field) power transported by the flow.  A first result of our analysis
(although strickly valid only in the case of IC/CMB sources, for which
all the quantities can be univocally evaluated) is that the power derived
at subpc scale is rather close to the power inferred at kpc scale. This
results (for the specific case of PKS 1510-089) is presented in Fig. 1a,
where we report the power estimated in the core region (horizontal dashed
line) and that at the kpc-scale jet expected for different $\Gamma $
factors, assuming that the emission is produced through IC/CMB (solid
line), through SSC (long-dashed line) and assuming equipartition between
relativistic electrons and magnetic field (short-dashed line). The
intersection between the equipartition and the IC/CMB lines marks the
$\Gamma $ factor used in the spectra modelling of the kpc-scale jet
emission. Note that for low values of Lorentz factors the power required
by the IC/CMB model increases rapidly, due to the large number of
electrons required to produce the observed X-ray flux in presence of a
rapidly decreasing amount of beamed CMB photons. Note also that an SSC
origin of the high energy radiation would require an extremely large
value of the power (exceeding $10^{50}$ erg/s), as early recognized for
the specific case of PKS~0637-752 (Schwartz et al. 2000).\\ The equality
of the jet power estimated at the two scales strongly suggests that the
jet experiences very little deceleration, dissipation and entrainment, at
least from the core until $\sim 100$ kpc. Recent numerical simulations
seems to confirm these conclusions (Scheck et al. 2002), at least for
distances $< 100$ kpc. Moreover the costancy of the power implies that
the central engine does not change its energy output for timescales of
the order of $\sim 3\times 10^5$ yrs.\\
\begin{figure}
\begin{tabular}{ll}
\hskip -0.3 truecm
\psfig{figure=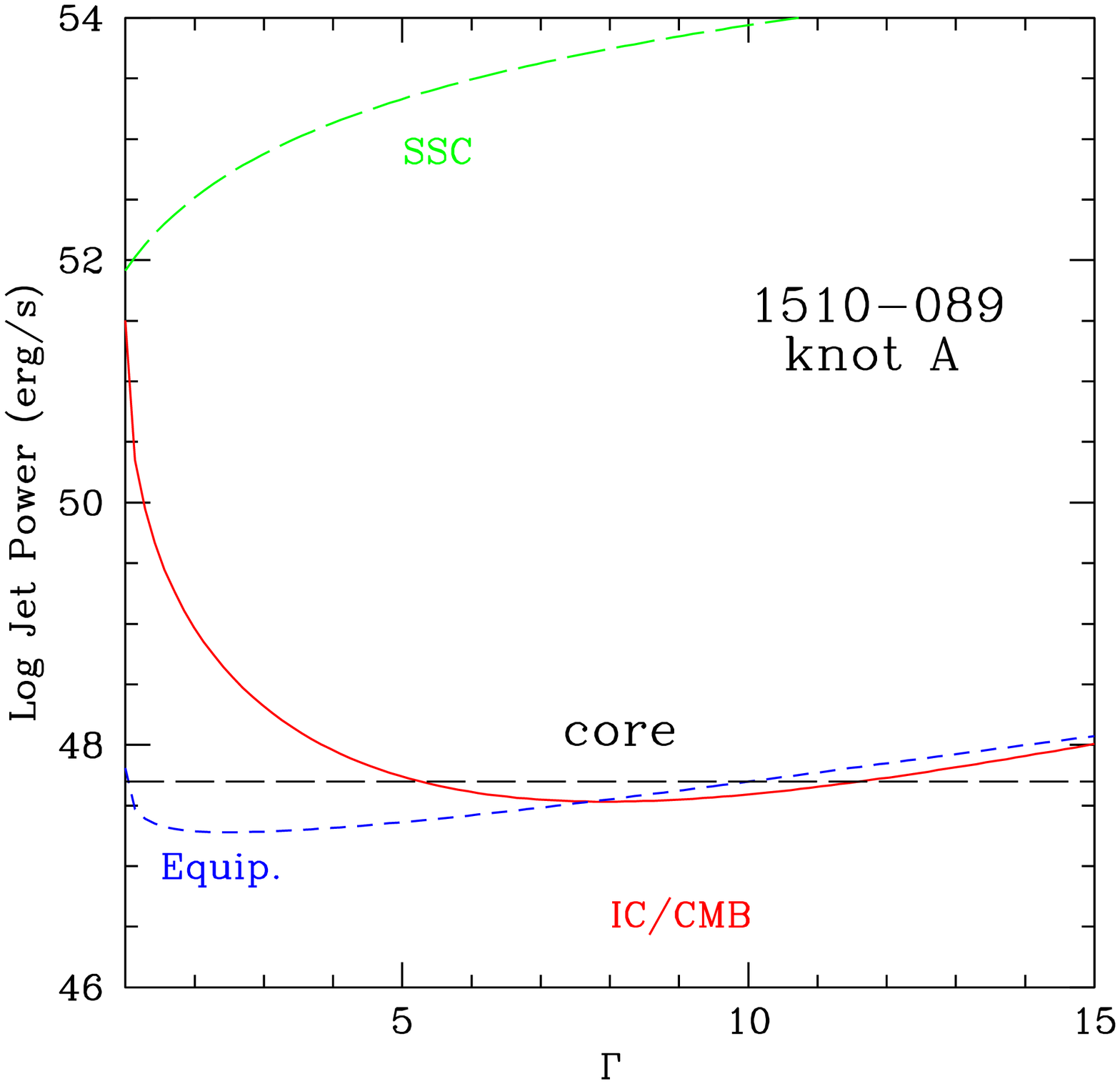,width=6.7cm}&
\psfig{figure=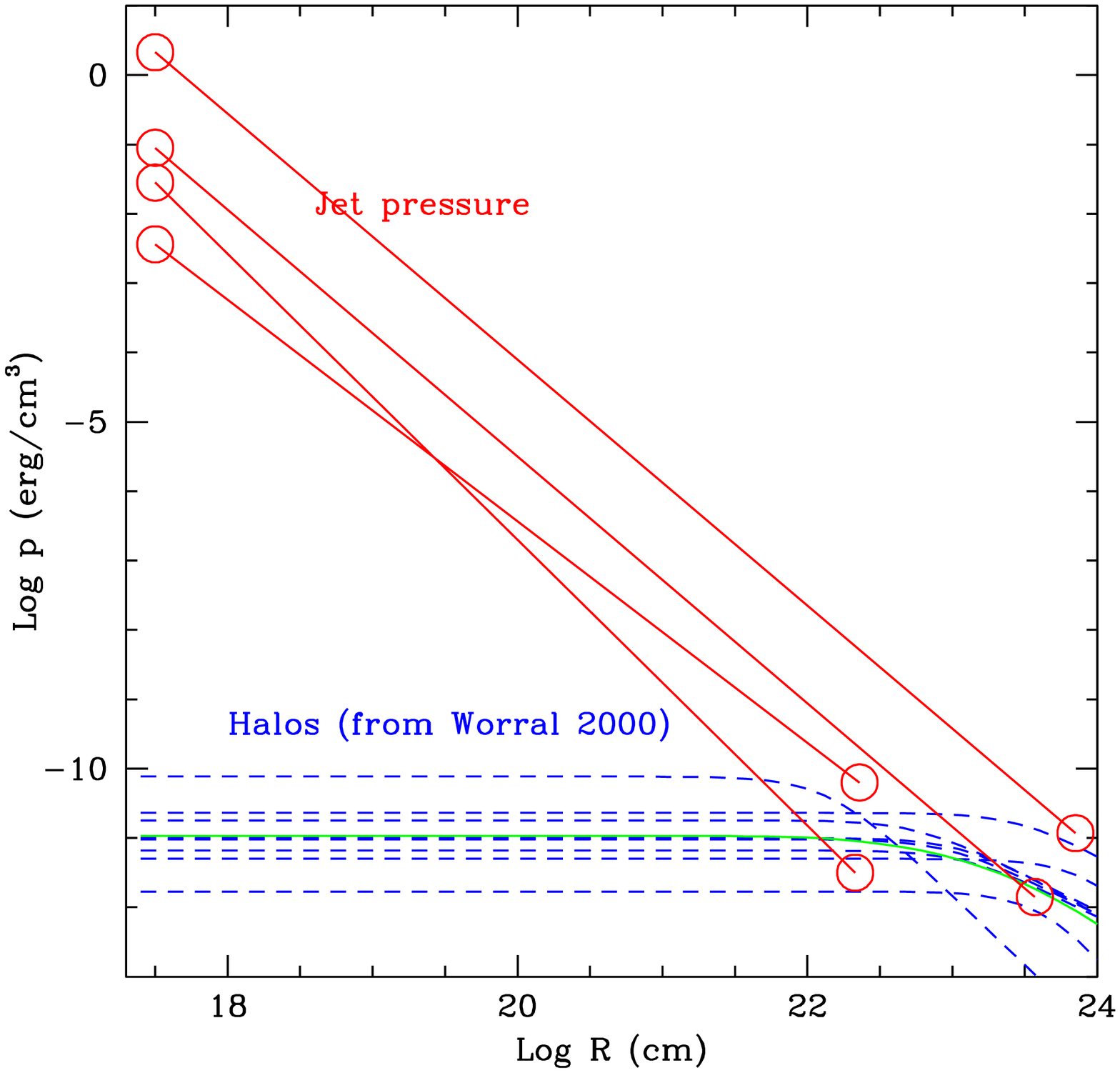,width=6.7cm}\\
\end{tabular}
\caption{{\it Left:} Jet power inferred at kpc scale and in the core for
1510-089. The power estimated using equipartition (short-dashed line) is
equal to that estimated through the IC/CMB (solid line) at $\Gamma \sim
7$ and it is very close to the core power. The power required by the SSC
model (long dashed) is much larger than the power usually estimated in
these sources. {\it Right:} Pressure profiles for the jet discussed in
the text (solid lines) and for the hot halos around a group of radio
galaxies (dashed lines from Worrall \& Birkinshaw 2000). The solid line
represents a logaritmic average of the profiles.}
\label{fig1}
\end{figure}
In our estimates we assumed that most of the power is carried by (cold)
protons (assuming the presence of 1 proton per relativistic
electron). While the composition of the outer jet is not known, in the
blazar region (see e.g. Tavecchio et al. 2000, Ghisellini \& Celotti
2002) the necessity of protons is clearly demonstrated by the fact that
the power transported by electrons and magnetic field alone are
insufficient to explain the total radiation output.\\ Another interesting
suggestion derives considering the internal pressure of the kpc-scale
knots inferred with our modelling. We recall that the estimate of the
pressure is rather rubust, due to the opportunity to strongly constrain
the value of the minimum Lorentz factor of the emitting electrons in the
IC/CMB model (Tavecchio et al. 2000a) The derived values are in fact
close to $p\sim 10^{-11}$ dyne cm$^{-2}$, suggestively close to the
pressure of typical hot gas halos found around FRIs (Worrall et al. 2000,
Worrall et al. 2001. Recently Worrall \& Birkinshaw (2000) reported an
analysis of a sample of B2 radio-galaxies with the HRI of ROSAT. Their
fits of the distribution with a King profile is reported in Fig. 1b, (we
also report the logaritmic average) together with our pressure estimates.
Notably, the expected external gas pressure is of the same order of the
pressure inferred in jets, strongly suggesting that the external gas and
the jet are in pressure equilibrium.

\section{Conclusions}

The results discussed above seems to point toward a simple scenario for
the dynamical evolution of the jet from inner to the outer
regions. Clearly the pressure in the blazar jet is larger than any
possible confining gas. This means that, after the dissipation region,
the jet expands freely until the internal pressure (gas and/or magnetic)
reaches the pressure of the (supposed) hot gas of the galactic halo. At
this point a reconfinement shock forms (the jet is highly supersonic),
particle are accelerated, and the jet starts to decelerate (Sanders 1983,
Komissarov 1994).  This scenario seems to account for the typical
distance at which knots form and also provide a reliable mechanism for
the production of relativistic particles. A even more interesting result
is that the expected pressure decay along the jet seems to be consistent
with the simple profiles predicted for a free, conical expanding jet
(Blandford \& Koenigl 1979).  A similar scenario has been recently
proposed to explain the dynamics of the jet in the FRI radio galaxy 3C31
by Laing \& Bridle (2002). The results discussed here are based on a very
small group of sources: future studies along these lines could further
support this scenario, providing important clues on the dynamics of
relativistic jets.




\end{document}